\documentclass[12pt]{article}

\usepackage{epsf,
color,amsmath,graphicx,amssymb,subfigure}

\newenvironment{sciabstract}{%
\begin{quote} \bf}
{\end{quote}}

\newcounter{lastnote}
\newenvironment{scilastnote}{%
\setcounter{lastnote}{\value{enumiv}}%
\addtocounter{lastnote}{+1}%
\begin{list}%
{\arabic{lastnote}.}
{\setlength{\leftmargin}{.22in}}
{\setlength{\labelsep}{.5em}}}
{\end{list}}

\title{Colloidal Jamming at Interfaces: a Route to Fluid-bicontinuous Gels}

\author{K. Stratford$^1$,  R. Adhikari$^2$, I. Pagonabarraga$^{2,3}$, \\
J.-C. Desplat$^{1,4}$, M. E. Cates$^{2\ast}$\\
\\
\normalsize{$^1$ EPCC, School of Physics, The University of
  Edinburgh,}\\ \normalsize{ JCMB Kings Buildings, Edinburgh, EH9 3JZ, UK}\\
\normalsize{$^2$ SUPA, School of Physics, The University of
  Edinburgh,}\\ \normalsize{ JCMB Kings Buildings, Edinburgh, EH9 3JZ, UK}\\
\normalsize{$^3$ Departament de F\'{\i}sica Fonamental, Universitat de Barcelona, }\\ \normalsize{ C. Mart\'{\i} i Franqu\'es 1, 08028 Barcelona, Spain}\\
\normalsize{$^4$ Irish Centre for High-End Computing,}\\
\normalsize{Dublin Institute for Advanced Studies,}\\
\normalsize{5 Merrion Square, Dublin 2, Ireland}\\
\\
\normalsize{$^\ast$To whom correspondence should be addressed; E-mail:  m.e.cates@ed.ac.uk}
\\
\normalsize{This is a preprint of Science v. 309 pp 2198-2201, published 30 Sept. 2005}
\\
\normalsize{
http://www.sciencemag.org/cgi/content/abstract/309/5744/2198}
}
\date{}

\begin{document} 


\baselineskip24pt


\maketitle


\begin{sciabstract}
Colloidal particles or nanoparticles, with equal affinity for two fluids, are known to adsorb irreversibly to the fluid-fluid interface.
We present large-scale computer simulations of the demixing of a binary solvent containing such particles. The newly formed interface sequesters the colloidal particles; as the interface coarsens, the particles are forced into close contact by interfacial tension. Coarsening is dramatically curtailed, and the jammed colloidal layer seemingly enters a glassy state, creating
a multiply connected, solid-like film in
three dimensions. The resulting gel contains percolating domains of both fluids, with possible uses
as, for example, a microreaction medium.
\end{sciabstract}

The search for new materials with mesoscale or nanoscale structure is a major theme of current physical science. Routes which exploit spontaneous self assembly in thermal equilibrium are certainly important \cite{whitesides1,whitesides2}, but nonequilibrium processes offer more control -- since assembly is then governed not just by thermodynamic conditions but by the entire process history (e.g. \cite{renth,blah,pinetemplate}). Moreover, the resulting materials may become trapped in deeply metastable states such as colloidal clusters, glasses, and gels \cite{renth,blah,pine,pham}, remaining more robust than an equilibrium phase to subsequent changes in thermodynamic conditions. This is a key consideration in determining the shelf-life and flow behavior of everyday products such as paint, vaccines, and yoghurt.

We use computer simulations to explore a kinetic pathway that leads to the creation of amorphous soft-solid materials. In such a material, which we call a `\underline{B}icontinuous \underline{i}nterfacially \underline{j}ammed \underline{e}mulsion ge\underline{l}' (or `Bijel', for short) \cite{patent}, a pair of interpenetrating, bicontinuous fluid domains are frozen into a permanent arrangement by a densely jammed monolayer of colloidal particles at the fluid-fluid interface. Such materials may have some remarkable properties, stemming directly from the nonequilibrium, arrested nature of the monolayer: Bijels should be highly tunable in elasticity and pore size through the volume fraction $\phi$ and radius $a$ of the solid particles. (The radius can be varied from microns to nanometres without altering the physics of structure formation by the route reported here.) One possible application, explored below, is as a cross-flow microreaction medium in which two immiscible fluids are continuously brought into intimate contact by pumping them in opposite directions through a static Bijel.

To achieve maximal stability of a particle-laden interface, the colloidal particles should be chosen with nearly equal affinity for the two liquids involved \cite{binksreview}. This creates similar values for the two fluid-solid interfacial tensions, and thus a fluid-fluid-solid contact angle close to 90 degrees (known as `neutral wetting'). A spherical particle is then in stable equilibrium with its equator positioned at the fluid-fluid interface. In practice, this equilibrium is so stable that detachment of such a particle cannot be achieved by thermal motion alone \cite{binksreview}. For neutral wetting, the fluid-solid interfaces have the same total energy regardless of particle position, but the fluid-fluid interfacial area is reduced, by a disc of radius $a$, when the particle lies midway across the interface. The detachment energy $\epsilon$ is the interfacial energy of this disc, $\epsilon = \sigma \pi a^2$, with $\sigma$ the fluid-fluid interfacial tension. Hence $\epsilon/k_BT = (a/a_0)^2$ where $a_0^2 = k_BT/\pi\sigma$. For $T = 300$ K and typical $\sigma$ of order 0.01 Nm$^{-1}$ or larger, $a_0$ is 0.4 nm or less.  Thus $\epsilon/k_BT \ge 10$ even for a particle of 1 nm radius, and thermally activated detachment can be safely neglected for, say, $a\ge 3$ nm.

We consider near-neutral wetting particles suspended in a binary solvent under conditions where the fluids are fully miscible (generally at high temperature) and of roughly equal volume fraction. In the absence of strong attractions between them, the particles will diffuse freely. If the temperature is quenched deep into a two phase region, the solvents will demix by spinodal decomposition \cite{onuki}. A sharp interface between the two fluids develops, and coarsens. During the coarsening, which is driven by the tendency of the interface to reduce its area, the characteristic lengthscale $L(t)$ initially increases with time in a well-understood manner \cite{kendon}, causing bumps on the interface to flatten, and causing necks between neighboring domains of the same fluid to pinch off. 

We have studied what happens after this initial separation by simulations \cite{methods} using the lattice Boltzmann (LB) method \cite{jsp,nguyen,desplat,ronojoy1,ronojoy2}. We find that, as coarsening proceeds, the interface sweeps through the fluid phases, efficiently collecting and sequestering the colloidal particles. Initially, the attached particles have little effect on the interfacial motion, but as more are collected and the interfacial area shrinks, they soon approach a densely packed monolayer. At this point, the fluid must either (i) stop coarsening at some lengthscale $L(t) = L^*$ or (ii) thereafter expel particles steadily from the ever-shrinking interface. In our simulations, we see a drastic curtailment of the coarsening and very little particle expulsion. This suggests that the free energy landscape of the dense colloidal film is such as to trap particles on the interface in a metastable, amorphous state  \cite{ergodic}.

Fig.1 shows the structure as time evolves; for visibility, only a small part (in cross-section, upper panels, or crop, lower panels) of the full system is shown. Ref.\cite{S1} shows an animation of the observed sequence of events within the cropped region of Fig. 1; 
this clearly shows the particle sweep-up and the dramatic slow-down of coarsening.
Ref.\cite{S2}
shows the entire simulation domain for 
another run, in which bidisperse particles are used; this prevents development of any local crystalline order in the interfacial colloid layer (arguably visible in Fig.1). Ref.\cite{S3} shows a similar subregion for this case as in Ref.\cite{S1}. The results are qualitatively the same: see Fig.2 for a snapshot at late times. Fig.3 shows the time dependence of the domain size $L(t)$ in each case.

The parameter values chosen for these simulations  \cite{methods} map onto particles of radius $a=5$ nm in a symmetric pair of fluids each having viscosity $\eta = 10^{-3}$ Pa s and mass density $\rho = 10^3$ kg m$^{-3}$ , with $\sigma = 6\times 10^{-2}$ Nm$^{-1}$ at $T = 300$ K; such values are typical of a short-chain hydrocarbon/water or alcohol/water mixture. Our particles have purely repulsive interactions, with range extending somewhat beyond their hard-sphere (hydrodynamic) radii \cite{methods}, so that particles remain visibly separated even in a dense monolayer. The parameter mapping is made by matching dimensionless control groups $\epsilon/kT$ and $a\rho\sigma/\eta^2$ \cite{codef,methods}. Brownian motion of the colloidal particles is included \cite{methods}, but has rather little effect during the time regime we can reach by simulation (see below), and would have even less effect with larger particles. We have also checked the role of short-range, thermally reversible bonding among colloids \cite{pham}, but this too has little effect. These observations are attributable to the strong separation between Brownian and interfacial energy scales referred to above.

A number of numerical compromises were made to keep the simulations tractable \cite{methods}. First,
the Reynolds number Re $=(dL/dt)\rho a/\eta$ is much larger than in the real system,  though we still have Re $\ll 1$ \cite{codef}. Second, the scale separation between the particle radius $a$ and the fluid-fluid interfacial thickness $\xi$ is only modest (a factor two or three), with the lattice spacing, in turn, not much less than $\xi$.  This gives imperfect discretization of the fluid-fluid interface around particles and may under-represent the energy barrier to short-scale rearrangements \cite{methods}. Finally, for the physical parameters given above, the
effective run-time of our largest simulations is about 300 ns. (For larger particles, say $a = 3\mu$m, the equivalent run time would be around 5 ms.) 

Although these simulations dramatically confirm our proposed kinetic pathway for creating a fluid-bicontinuous state with a particle-laden interface, they cannot tell us whether this state is a fully arrested gel on laboratory time scales.
Unlike our simulations, the latter vastly exceed the time scale $\tau = 6\pi\eta a^3/k_BT$ characterising Brownian motion of colloids. Nonetheless, the observed behavior, particularly for bidisperse particles, is consistent with that of a long-lived, metastable, arrested state. In common with other such states (including colloidal glasses), Bijels might show slow aging behavior as the interface approaches saturation. This may explain the residual late-time dynamics visible in the $L(t)$ curves of Fig.3.
Alongside aging, the slow residual dynamics could be due to the incomplete separation of length scales in LB noted above or, in the monodisperse case, due to a tendency for the interfacial layer slowly to acquire local crystalline order. (Such ordering would not preclude, and might even enhance, eventual structural arrest.) 
We have also assessed the particle mobility in the interfacial film by measuring the distribution of individual particle displacements at late times. We found this to be dominated by the residual relaxation dynamics of the structure rather than by diffusion within the film. 

The near-complete cessation of fluid motion on time scales of order $\tau$ suggests that further coarsening, which requires expulsion of particles from the interface, cannot take place without colloidal Brownian motion. If, as argued earlier, such motion is ineffective at detaching particles from a static interface, coarsening must cease altogether. In principle, lateral diffusion within a film of fixed area might continue. However, the surface pressure of such a film is estimated \cite{schowalter} as $\Pi \sim k_BT/[ a^{2} (\psi_m-\psi)]$ where $\psi$ is the areal fraction and $\psi_m$ that of a maximally close-packed configuration. For this pressure to balance the interfacial tension requires $\psi_m-\psi \sim (a_0/a)^2$. For large enough particles, this
ensures that $\psi$ exceeds any threshold value, so long as this value is less than $\psi_m$, for the onset of an arrested state within the film.

Further evidence for interfacial arrest was gained by additional, higher-resolution simulations that examine the dynamics of two specific structural motifs characteristic of the bicontinuous structure. One of these is a long cylinder, representing a fluid neck. Without particles the Rayleigh-Plateau instability \cite{siggia} would cause the cylinder to break into droplets. We show in Fig.S1 a dense bidisperse colloidal packing on a cylindrical interface. When perturbed, this shows no sign of either ordering or breakage, and the initial undulation visibly decays, rather than grows. The structure then arrests, and remains unchanged for the duration of the simulation, which is ten times longer than the time-to-rupture of an unprotected cylinder.
Our second structural motif is a periodically rippled surface, whose bumps are broadly characteristic of a non-necklike section of the
bicontinuous interface between fluids. Without particles, the ripple would rapidly be pulled flat by interfacial tension.
Fig.S2 shows how this process is interrupted by interfacial jamming. After an initial transient decay, the film jams, and bumps on it persist at least thirty times longer than the decay time without particles. There is negligible macroscopic motion during the latter half of this simulation. 

These results show that, at sufficient interfacial coverage \cite{diskpack}, both necks and bumps can arrest by jamming of the adsorbed colloidal layer into a densely packed, and therefore presumably glass-like, state. Since these two structural elements are, in combination, the driving features of bicontinuous coarsening \cite{siggia}, their arrest would be enough to prevent coarsening. Hence these studies provide very strong supporting evidence for an eventual arrest of our bicontinuous structure, caused by a jamming transition \cite{jamming} within the colloidal monolayer. This transition is induced by the (capillary) stresses arising from the fluid-fluid tension.

Once the interfacial film has fully arrested, since it percolates in three dimensions, the entire material will acquire solid elasticity at scales beyond $L^*$. The static modulus $G$ of the resulting gel should scale with the interfacial energy density $\sigma/L$; so long as nearly all particles end up on the interface, $L^*\sim a/\phi$ with $\phi$ the particle volume fraction. For $\sigma = 0.01$ Nm$^{-1}$, $0.01\le \phi \le 0.1$ and $5$nm $<a < 5\mu$m we estimate $20\le G\le 2\times 10^5$ Pa. This is a very wide `tuning' range for material design. Under nonlinear stress the interfacial area will dilate significantly: only a modest dilation (say 10\%) may suffice to cause melting of the particle layer and drastic fluidisation. This might instigate both flow and coarsening above some yield stress $Y\simeq 0.1 G$. If the stress falls back below $Y$, we expect resolidification, possibly with remanent anisotropy (hysteresis). The nonlinear flow behavior of these gels could thus show a remarkable strain-melting, possibly reminsicent of a colloidal glass, but with a much higher stress scale set by interfacial, not Brownian, forces. 
The estimates above for the material properties of the gel stem from the jamming of colloids by the interfacial forces and apply even for purely repulsive particles. 
Any additional bonding attraction, if of sufficient strength, might enhance the rigidity of the interfacial layer, but may also cause colloidal aggregation within the bulk phase(s) prior to monolayer formation \cite{clegg}. Fusing the colloids after gel formation (e.g. by irradiation) would completely stabilize the structure and drastically alter the flow behavior.

Our study differs from previous work in which colloidal particles were used to stabilize spherical
\cite{binksreview,weitz} and aspherical \cite{clegg} emulsion droplets of one liquid in another. Under compression, such emulsions can form robust gel phases with interesting mechanical properties \cite{arditti}, but fluid bicontinuity is not among them. The previously preferred route to particle-stabilised emulsions involves agitation of immiscible fluids and does not appear to favour bicontinuity \cite{binksreview}. Other related work \cite{pine,maher2,balazs,nematics} involves particles with a strong preference for one of the two liquids, creating a particle network within the chosen liquid rather than at the interface.

We expect Bijels to have several interesting physical properties beyond those discussed above. First, the fluid-bicontinuous state should remain equally insoluble on exposure to either of its solvents. This contrasts with particle-stabilized emulsion gels formed by compression \cite{arditti}, in which an excess of the continuous phase could cause droplets to separate, losing macroscopic rigidity. In Bijels this will not happen, since neither of the two interpenetrating fluids can alter its volume without also  increasing the total interfacial area. The Bijel can thus metastably support simultaneous coexistence with bulk phases of both fluids. This is reminiscent of an equilibrium property of middle-phase microemulsions, which are bicontinuous states with an interface stabilized by amphiphilic molecules \cite{microemulsions}. In contrast to Bijels, such microemulsions are not gel phases, but inviscid fluids, as a result of their high interfacial mobility. 

Second, fluid bicontinuity imparts high permeability of the gel to either of its component solvents, and any reagents dissolved in them. Accordingly Bijels may have potential as media for continuous-process microreactions \cite{robinson1,robinson2}. Specifically, a static gel could support a steady permeation flow of both fluids simultaneously in opposite directions. This would bring two molecular reagents, soluble only in mutually immiscible fluids, into intimate contact at the fluid-fluid interface in the interstitial regions between the colloids. Reaction products soluble in either phase would be swept out continuously.
To test this concept, Fig.4 shows a simulation in which the two fluids are moving through the structure in opposite directions. On the timescale of our simulation, the gel has easily enough mechanical integrity to sustain this cross-flow without breaking up. Within the mapping onto physical parameters made previously, the chosen cross-flow fluid velocity $v= 0.01\sigma/\eta$ is of order 10 cm s$^{-1}$: this is an extremely large value, given the pore scale $L^*$ of only 70 nm. Local shear rates are of order $10^6$ s$^{-1}$.

In summary, we have presented simulation data showing formation of a self-assembled bicontinuous structure with interfacially sequestered particles. This followed a kinetic pathway involving a colloidal suspension in a binary solvent, initially miscible, that undergoes a temperature quench. Our simulations show a drastic curtailment of coarsening, consistent with eventual structural arrest: a scenario further supported by higher-resolution studies of appropriate structural motifs (bumps and necks). This suggests a route to the creation of  new class of gels, Bijels, with potentially remarkable physical properties.

\begin{scilastnote}\item

We thank Paul Clegg, Stefan Egelhaaf, Eva Herzig, Wilson Poon and Bernie Binks for discussing with us their ongoing experiments on colloidal jamming at interfaces \cite{clegg}. We thank Eric Weeks for advice on visualisation, A. Donev for supplying code for the algorithm of \cite{diskpack}, and Claire Rutherford for discussions. This work was supported by EPSRC through Grants GR/S10377 and GR/R67699 (RealityGrid). IP acknowledges financial support from DGICYT of the Spanish Government and from DURSI, Generalitat de Catalunya (Spain). 

\end{scilastnote}

\clearpage

\begin{figure}
\color{white}
\begin{center}
\includegraphics[width=140mm]{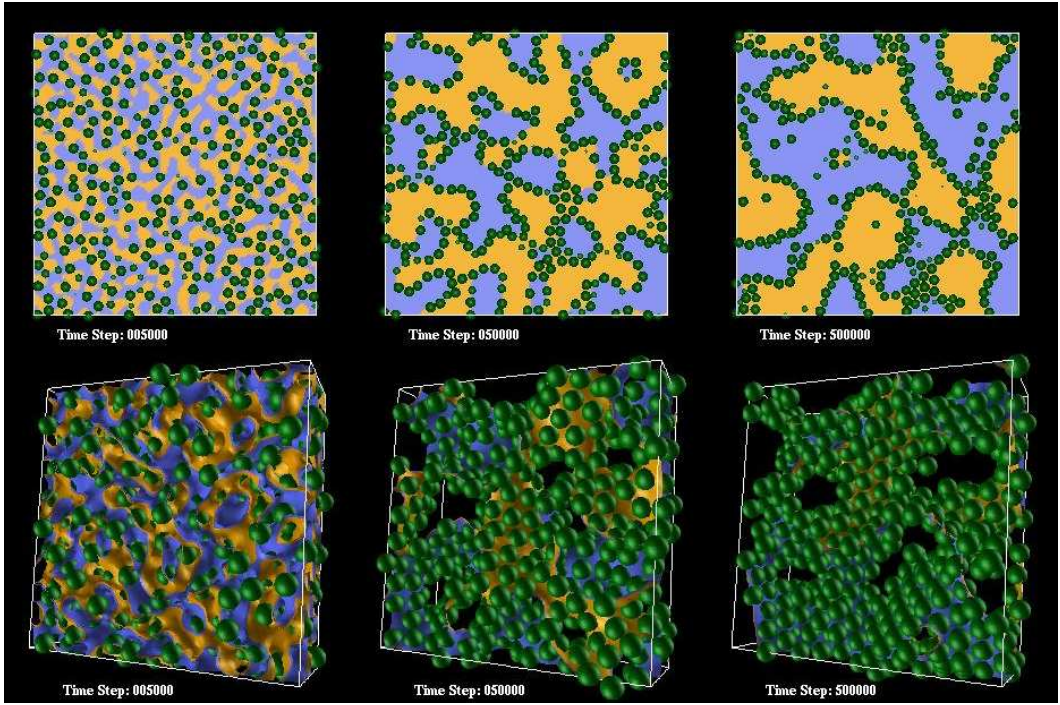}
\caption{
\color{black}
{\bf Fig.1.}
Time evolution of monodisperse neutrally wetting colloidal particles at volume fraction $\phi =0.2$ in a binary solvent following a quench. In the upper panels, a $128\times 128$ section through the full simulation (performed on $128^3$ lattice) is shown. The two fluids are colored yellow and blue. Monodisperse colloidal particles (green) are shown only if overlapping the plane of the section, with parts lying behind this plane occluded (so that particles whose midpoint is behind the plane appear reduced in size). Sequestration of the particles, followed by near-arrest of the bicontinuous structure, is seen. Frames are at 5 000, 50 000, 500 000 LB timesteps; for parameter details see \cite{methods}. In the lower panels
we show a three dimensional view of a $64\times 64\times 24$ piece cropped from the same simulation with the same time-frames. Both fluids are now shown transparent with the two sides of the fluid-fluid interface painted yellow and blue. Sequestration, arrest and fluid bicontinuity of the resulting structure are all apparent. 
}
\end{center}
\end{figure}

\clearpage

\begin{figure}
\color{white}
\begin{center}
\includegraphics[width=140mm]{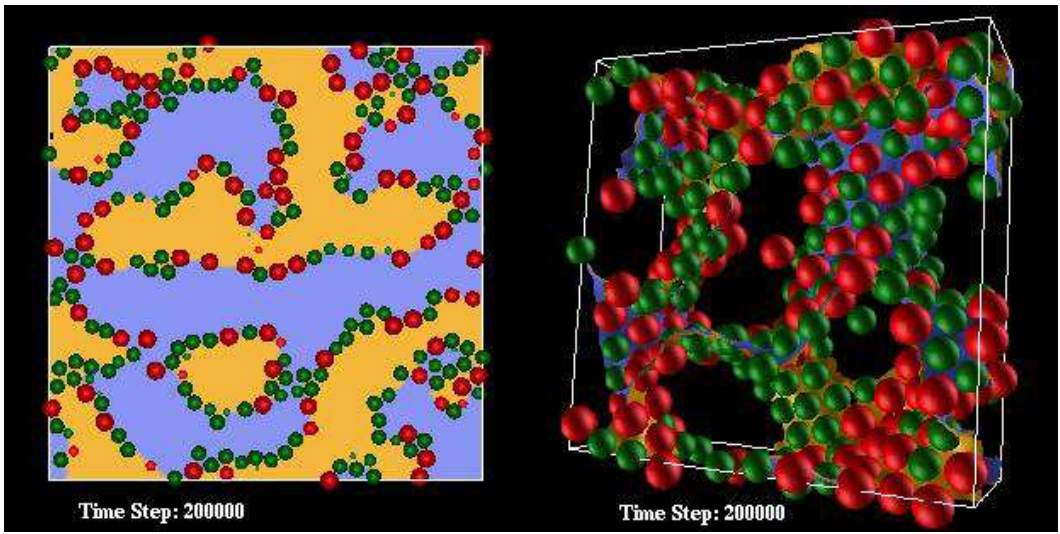}
\caption{
\color{black}
{\bf Fig.2.}
Structure at 200 000 timesteps in a simulation of bidisperse neutrally wetting particles in a binary solvent following a quench, in section and 3D view. Color coding as in Fig.1 but with larger particles shown in red; size ratio 1.2:1. On the left is a $128\times 128$ section;
the top left quadrant of this square is coincident with the front face of the $64\times 64\times 24$ cuboid used for the 3D view (at right). There is no sign of local crystalline order in the particle film (visible in Fig.1, lower panels) although some tendency toward local segregation by particle size is seen. For parameter details see \cite{methods}. The same structure, evolved to 590 000 timesteps, is shown in Fig.4 below.
}
\end{center}
\end{figure}

\clearpage

\begin{figure}
\color{white}
\begin{center}
\includegraphics[width=140mm]{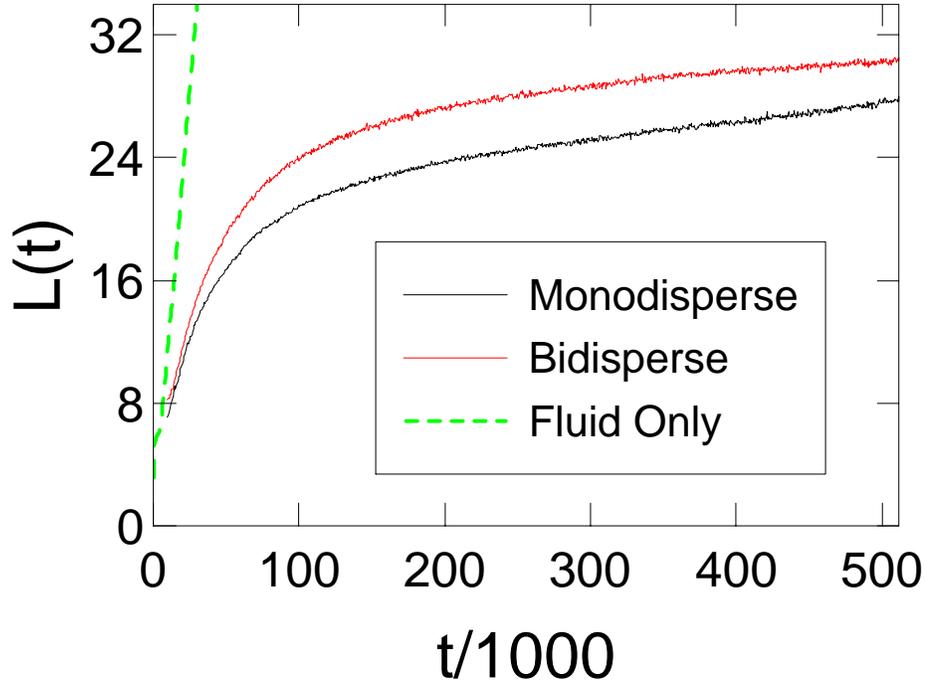}
\caption{
\color{black}
{\bf Fig.3.}
Time evolution of the structural length scale $L(t)$ for monodisperse (lower curve) and bidisperse (upper) particle runs as described in \cite{methods}. $L$ is measured in units of the LB lattice and $t$ in LB timesteps. Without particles, coarsening would proceed with the slope indicated by the dotted line (see \cite{kendon}). Near-arrest is visible, suggesting a finite asymptotic domain size $L^*$, particularly in the bidisperse case. This $L^*$ is less than 1/4 of the simulation box size and not limited by finite-size effects \cite{kendon}. For the parameter values chosen (corresponding to 5nm particles) the data shown run approximately from 6 ns to 300 ns in real time \cite{methods}, with $L^* \simeq 70$ nm. (At times less than 6 ns, the fluids are demixing diffusively so that sharp interfaces have not yet formed.)
}
\end{center}
\end{figure}

\clearpage

\begin{figure}
\color{white}
\begin{center}
\includegraphics[width=80mm]{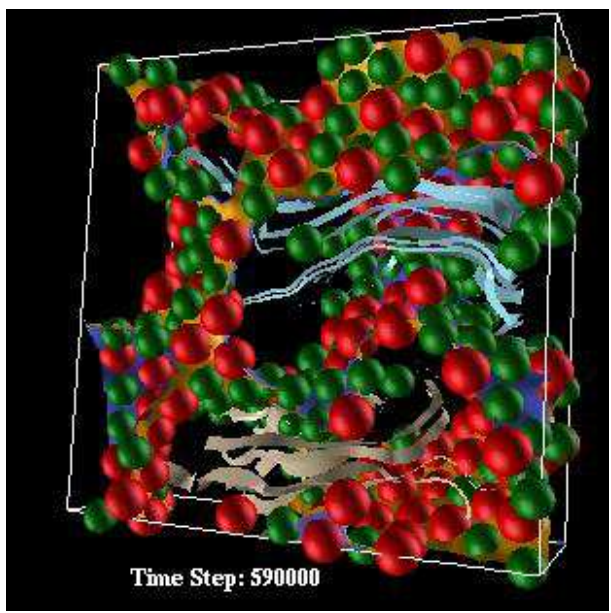}
\caption{
\color{black}
{\bf Fig.4.}
A $64^2\times 24$ section of the near-arrested bicontinuous structure in the run of Fig.2, showing velocity streamlines for the two component fluids under cross-flow forcing with velocity of order $0.01\sigma/\eta$ \cite{methods}. 
The group of beige-colored
steamlines starting at the bottom left show flow in one fluid (the yellow side of the interface)
moving rightward; the second group (blue) starting on the upper right show
flow in the second fluid moving to the left.
There is no discernible motion of the interfacial structure at this flow rate.
}
\end{center}
\end{figure}

\clearpage




\topmargin 0.0cm
\oddsidemargin 0.2cm
\textwidth 16cm 
\textheight 21cm
\footskip 1.0cm

\renewcommand{\vec}[1]{\ensuremath{\mathbf{#1}}}
\renewcommand{\psi}{\varphi}

\section*{Appendix: Supporting Material}

\subsection*{Methods}

In the simulations reported here, we choose for simplicity a perfectly
symmetric pair of fluids with equal density $\rho$ and viscosity $\eta$.
The phase diagram that controls their demixing is also symmetric, being
described by the free energy functional {\em (S1)}
\begin{equation}
F[\psi] = A\psi^2/2 + B\psi^4/4 + \kappa (\nabla\psi)^2/2
\end{equation}
where the order parameter $\psi$ describes the fluid composition, and
the choice of the parameters $A$, $B$, and $\kappa$ controls the
fluid-fluid interfacial tension $\sigma$ and thickness $\xi$ 
{\em (S1)}.

The two
solid-fluid interfacial tensions are exactly equal, and the interfacial thermodynamics implemented as in 
{\em (S2)}.
We choose a
deep quench in which the fluid-fluid interface should be sharp
on the scale of a colloid (see below) and the thermal excitations
of the interface (thermal capillary waves) are negligible. Thermal noise is however fully
included in the description of fluid momentum --- this imparts
Brownian motion to the colloids. Physically relevant control
parameters are then the viscosity $\eta$, the particle radius $a$,
the fluid-fluid interfacial tension $\sigma$ and the thermal
energy $k_BT$. (The fluid density $\rho$ can be scaled out.)

\if{The simulations were designed with a short-chain hydrocarbon/water or hydrocarbon/alcohol mixture at room
temperature in mind, for which typical parameters are
$\rho = 1000$~kg~m$^{-3}$, $\eta = 10^{-3}$~N~m$^{-3}$ (both
appropriate for water at 300K) and $\sigma = 6\times 10^{-2}$N~m$^{-1}$. 
The colloid radius was chosen to be in the nanometre range; in the simulation this controls the importance of Brownian motion relative to interfacial forces. However
additional simulations (not shown) with noise levels chosen to model colloid radii in the micrometre
range show very similar results, both in the sweepup stage and the later stage of dramatically inhibited coarsening.}\fi

For our binary fluid system we use the lattice
Boltzmann method, incorporating spherical solid particles 
{\em (S2)}.
Our code currently runs under OpenMP  on shared memory parallel machines, while an MPI (Message Passing Interface) version is currently under development for use on larger, distributed memory machines.
For the binary fluid alone, spinodal decomposition has been extensively
studied with this code 
{\em (S1)}
and our methodology is well validated in that context. For colloids we use a standard
`bounce-back on links' method 
{\em (S3)},
modified to allow
for the binary solvent 
{\em (S2,S4)}.
Brownian motion is achieved by a fluctuating fluid method distinct from, but closely related to, that of Ladd 
{\em (S5,S6)}.

The
characteristic length and time scales associated with the physics of coarsening are 
{\em (S1,S7)}
$L_0 = \eta^2 / (\rho \sigma)$ and
$t_0 = \eta^3 / (\rho \sigma^2)$, which for the physical parameters chosen in the main text
are $L_0 \approx 14$~nm and $t_0 = 0.22$~ns.
Computing the same quantities in the `lattice units' of LB 
{\em (S1)}
allows length and time scales to be matched to experiment, in principle. However, not all the dimensionless control parameters can fully be matched in practice. For example the Reynolds number, $Re = (dL/dt)\rho a/\eta$, which characterises the relative importance of fluid inertia to viscosity, can be made small compared to unity, but not as small as the physical value 
{\em (S8)}.
We fully match $\epsilon/k_BT$ and $a/L_0$ as described in the main text. The identification of our longest runs as 300 ns in duration then follows from the definition of $t_0$. To represent fluid-fluid-solid wetting behavior fauthfully we also require $1\ll\xi\ll a$ in lattice units; but this is only marginally achievable (see below for values).

We performed two simulations using a lattice
of 128$\times$128$\times$128 sites with periodic boundary conditions 
{\em (S1)}.
The main production runs took around 1 week on a 32-processor IBM p690+ system and some 2-3 weeks on
a 48-processor Sun E15K system. These were accompanined by further runs at the same scale, and many smaller $64^3$ and $32^3$ runs, to check
that physical trends were as expected.
If the
system is too small, fluid motion will be artificially arrested once $L$ approaches the box size (with the
interface then attaining a state of zero mean curvature in three dimensions) even in the absence of a monolayer of particles 
{\em (S1)}.
This state was often reached with our smaller system sizes.
However, Fig.3, which shows the domain length scale as a function of time
(computed following 
{\em (S1)}),
confirms that in our 128$^3$ runs, $L$
remains  significantly less than the system size: the drastic slowdown of coarsening is not a finite-size effect.

Our free energy
parameters were $A=-0.002$, $B=0.002$, and $\kappa = 0.0014$ giving an
interfacial thickness of $\xi = 1.14$, tension
$\sigma = 0.0016$, fluid density $\rho = 1$, and
viscosity $\eta= 0.1$ (all in lattice units). 
The fluid was initialised to be well mixed and at rest. A small amplitude
random noise was added to the $\psi$ field to
induce spinodal decomposition. At the same time, colloids were
positioned at rest randomly throughout the system. 
Thermal fluctuations appropriate
to a temperature of 300~K were included 
{\em (S5)}.
The first simulation (Ref.\cite{S1}) is a monodisperse suspension with 8229 particles
of radius $a = 2.3$ lattice units (corresponding to $5.4$ nm in physical units) providing a solid volume fraction
of 20\%. The bidisperse simulation (Refs.\cite{S2,S3}) has 4114 colloids of radius $a=2.3$
and 2407 larger particles of radius $a = 2.74$ lattice units. The overall solid volume fraction is again 20\%.
\if{(Note that
the total area $N\pi a^2$ is slightly greater in the first case, so
the final domain size is smaller then in the second case (Fig.3).)}\fi
Both simulations were run initially for 520,000 
time steps, which is 275~ns in physical time; the bidisperse run was then extended, to examine cross-flow (see below). 
 
Note that longer physical time scales (of order milliseconds) would be achieved if we chose parameters to model micron-scale colloids rather than nanocolloids. However, it is currently not practicable to run for timescales very long compared to the Brownian relaxation time of a free colloid (of any size) while still maintaining realistic values for $L_0$ and $t_0$ as required for the coarsening problem 
{\em (S8)}.
On the other hand, the Brownian timescale can be reached by artificially raising the temperature; this was done for the structural motifs
(cylinder and ripple) as described below.

On the lattice the colloids are discrete, block-like, objects. To
take account of this, a calibration is performed to compute an
appropriate hydrodynamic radius $a_h$ 
{\em (S3)}.
This
is the radius of the sphere which exhibits the same mean Stokes drag factor
$6\pi\eta a_h$ as the discrete colloid on the lattice. For the viscosity used here
($\eta = 0.1$), the actual and hydrodynamic radius for the smaller
colloids are found to be the same, $a = a_h = 2.3$, while for the
larger particles the hydrodynamic radius is slightly larger ($a = 2.74$
and $a_h = 2.75$). Fluid-mediated interactions between the
particles are well represented within LB when colloids are separated on the lattice, but this
breaks down when the colloid-colloid separation $h$ drops below the lattice
scale. This can be rectified by a standard procedure in which lubrication forces are added by hand 
{\em (S2,S3)}.
In our runs, the normal ($h^{-1}$) component of the pairwise lubrication
interaction is corrected at interparticle separations $h < h^* = 0.7$
lattice units. A much weaker transverse component of the lubrication force is neglected 
{\em (S2)}.

The computation of the lubrication forces itself becomes a major numerical exercise, with bad ($N^3$) scaling in the number $N$ of colloidal particles in simultaneous mutual lubrication contact 
{\em (S2)}.
Since sequestration at the fluid-fluid interface results in very large $N$, a workround for this is essential. We achieve it by adding an
additional pairwise thermodynamic potential ($\propto h^{-2}$) which effectively prevents particles approaching closer than
roughly 0.3 lattice unit. This results in a visible residual
spacing between particles in the interfacial monolayer (see figures).
Such short range repulsions are quite common physically 
{\em (S9)}
and
do not seriously compromise the realism of our simulations.

In contrast to this treatment of the lubrication forces, no equivalent corrections to the interparticle forces take place in
the thermodynamic sector. Thus, for a dense particle layer, there may be relatively few fluid-fluid nodes left in the interstices. This underestimates interfacial energies, and could account for continued slow coarsening and detachment after particles become densely packed on the
interface (Fig.3). In particular, a narrow neck, only one or two particles across, could become internally `dry' with no fluid nodes containing the enclosed solvent.
This problem is alleviated for our higher resolution runs for structural motifs (Figs. S1, S2); these have somewhat larger particle size and considerably larger length-scales for the interfacial structures (necks, bumps) themselves.

For the cross-flow simulation (Fig.4), a $\psi$-dependent body force was applied to the fluid which drives the different phases in opposite directions. A pumped flow (driven by pressure gradients) is not expected to differ significantly. The body force was switched on after near-arrest was complete at 520 000 time steps and the simulation run on to 600 000 steps to allow a near-steady flow to establish. The mean of the velocity was recorded over
the final 20 000 steps and used to generate the streamline ribbons
seen in Fig.4. A small number
of free particles (not attached to the interface) move discernibly during this time; the interfacial motion itself is negligible (the image shows the
particle and interfacial configuration at 590 000 time steps).
In a separate run (not shown) we increased the forcing to test the resilience of the structure. A transient ($\sim 0.5$ ns) forcing twenty times stronger than in Fig.4 led to significant distortion of the interfacial structure followed by partial elastic recovery when forcing was removed. However, it was apparent that this forcing, if maintained, would lead to structural meltdown.

To check the role of Brownian motion, we switched this off midway through a run whose initial parameters were that of Ref.\cite{S1}. There was a reduction in visible wobbling of particles at the interface, but little effect on the macroscopic motion. Reducing the thermal noise level is equivalent to increasing the particle radius; even with no noise from the outset of the run, we found very similar results to those presented above. For the structural motif work, we also tried artificially increasing the temperature 16-fold part-way through the cylinder and sinewave runs. Enhanced wobbling of particles was seen, but no major structural relaxation occurred on the Brownian timescale of the particles. This suggests that the physics of arrest is largely independent of Brownian motion and hence of particle size. It remains possible that structural relaxation takes place on a timescale well beyond the Brownian
time $\tau = 6\pi\eta a^2/k_BT$, and thus not probed in our simulations. However, since the interface has come virtually to rest on this timescale, 
and the energy barriers to both particle ejection and particle rearrangement are set by $\sigma\pi a^2 \gg k_BT$, it seems unlikely that Brownian motion could overcome these barriers unaided.
Similar checks were made for the role of attractive bonding interactions between colloids. It is possible that very strong attractions (as might be required to compete effectively with interfacial forces) could have a strong effect, but for bonding energies of up to  several times $k_BT$ we found no discernible difference from the purely repulsive runs described above.

The higher-resolution studies (Figures S1,S2) were done using the same fluid parameters as before, but with somewhat larger average particle sizes (2.7 and 4.1 lattice units for the cylinder, 2.1 and 3.2 for the rippled surface). These particle sizes correspond to physical radii of 5 nm -- 9 nm for our model aqueous/hydrocarbon mixture ($\eta = 10^{-3}$ Pa s, $\rho = 10^3$ kg m$^{-3}$ and $\sigma = 6 \times 10^{-2}$ N m$^{-2}$ at 300 K). For these studies, the thermodynamic interaction potential used to maintain a nonzero surface-to-surface contact distance $h$ in the packed film comprised a screened Coulomb interaction with Debye length $\lambda = 0.2$ lattice units, truncated with offset so as to vanish at and beyond $h = 0.4$ lattice units. The amplitude of the interaction force (effectively, the surface charge 
{\em (S9)})
is chosen to engineer an equilibrium value of $h = 1.0$ lattice unit for a regular triangular lattice of particles of the harmonic mean size
as estimated by minimizing the total energy of a unit cell of this lattice. This maintains surface-to-surface spacing of particles in a dense layer at of order one lattice spacing, which, for the chosen particle sizes, ensures that the thermodynamics of the fluid-fluid interface in the interstices between particles is adequately resolved by the discretization.

\newpage

\begin{figure}
\color{white}
\begin{center}\subfigure[]{\includegraphics[%
  scale=0.23]{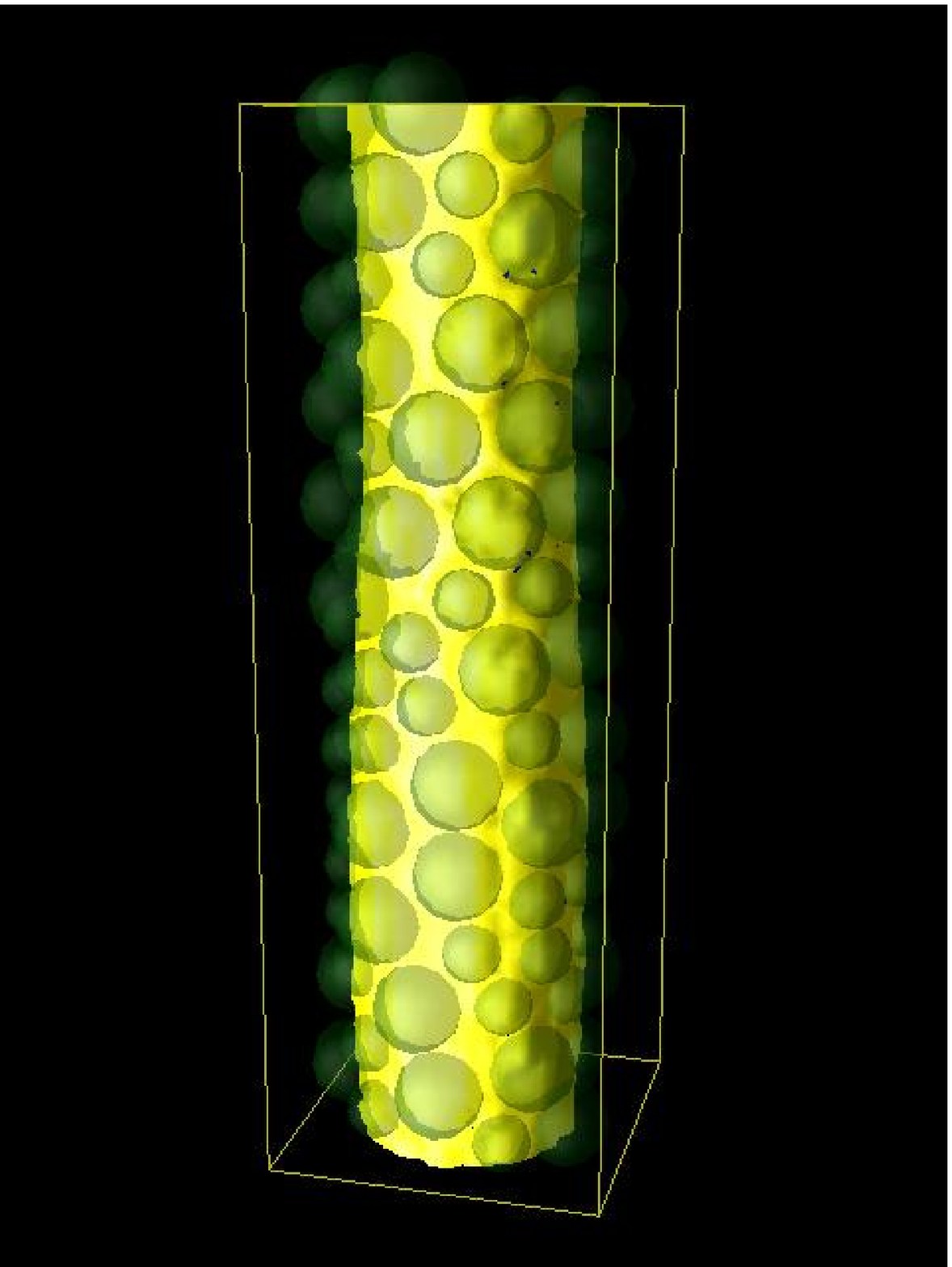}}\hskip0.05truein\subfigure[]{\includegraphics[%
  scale=0.23]{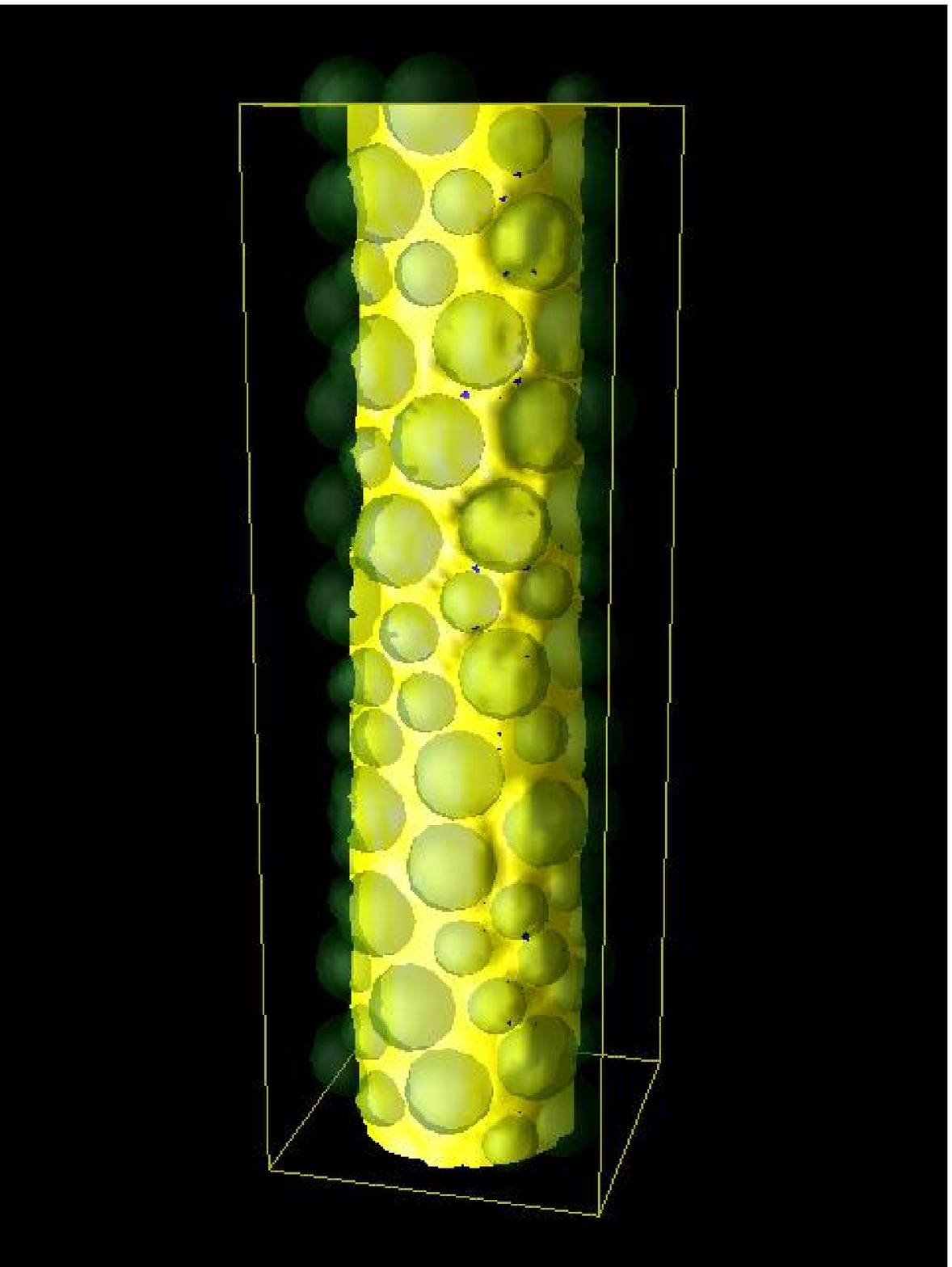}}\hskip0.05truein\subfigure[]{\includegraphics[%
  scale=0.23]{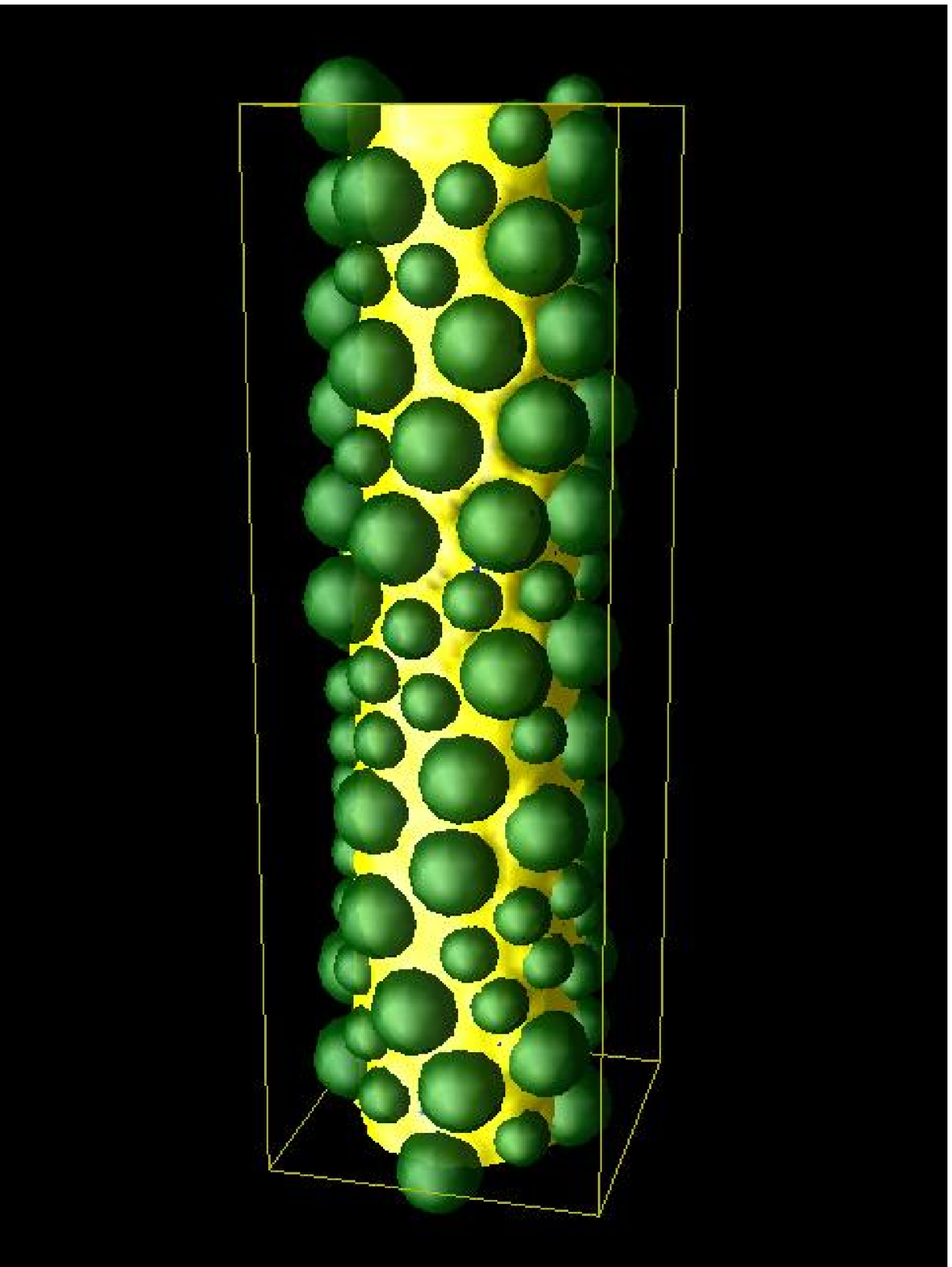}}\hskip0.05truein\subfigure[]{\includegraphics[%
  scale=0.23]{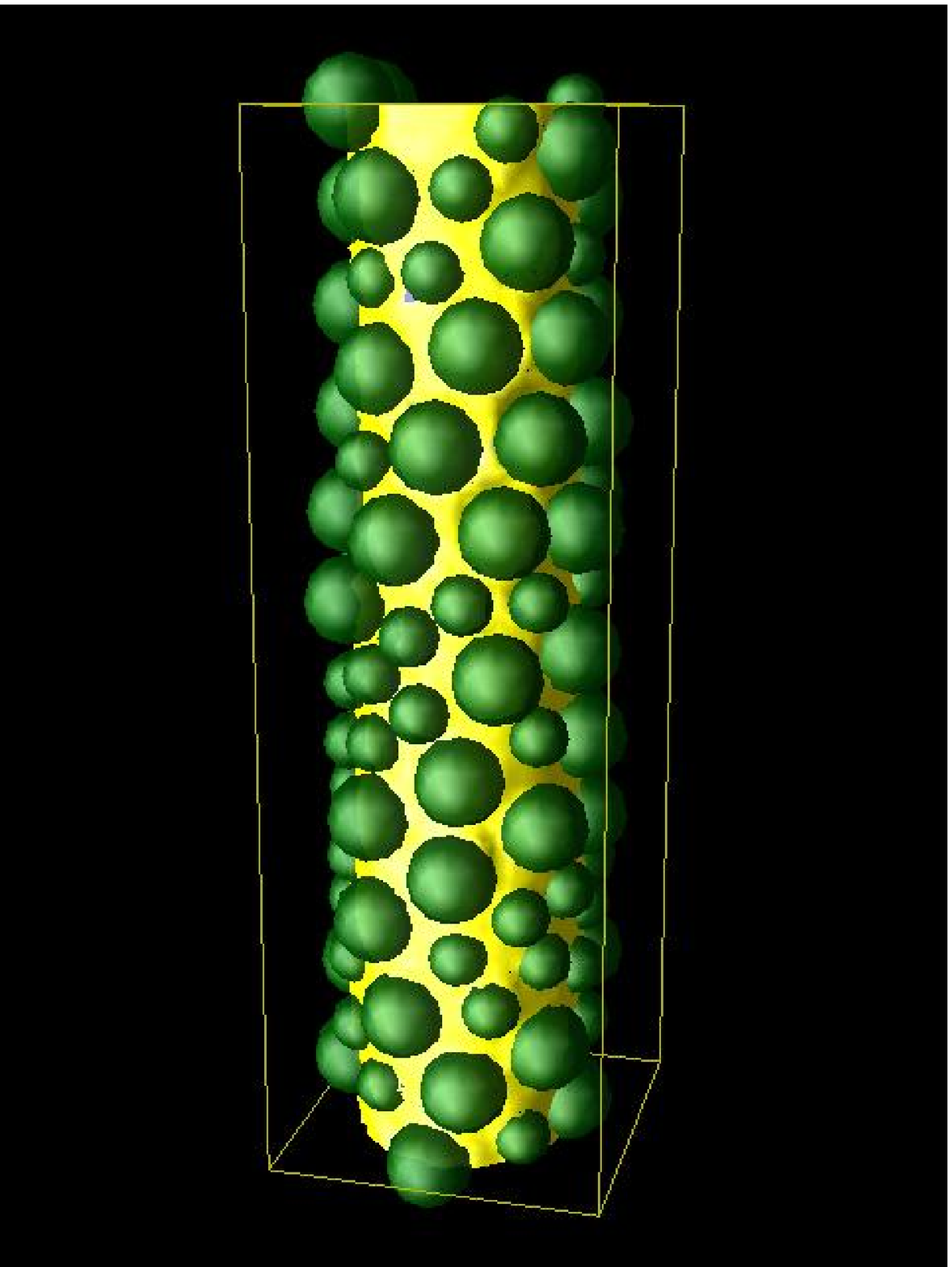}}

\caption{
\color{black}
{\bf Figure S1:} Time evolution of a cylinder coated with bidisperse, neutrally wetting colloids. Left frame (particles shown translucent) shows the perturbed interfacial configuration shortly after initiation (5 000 timesteps). Instead of growing (as occurs for the particle-free cylinder via the Rayleigh-Plateau instability, culminating in pinch-off), the perturbation decays to a smaller amplitude and then arrests: second frame, particles again translucent, after 100 000 timesteps; third frame, 200 000 timesteps. There is almost no visible evolution between this and 600 000 timesteps (rightmost frame); without particles, rupture occurs at $t = \tau_r\simeq$ 55 000 timesteps. The system occupies a $32\times 32\times 96$ lattice; particle radii are 2.7 and 4.1 lattice units. The initial state was created using a planar packing generated by the algorithm of 
{\em (S10)}
on a periodic domain with coordinates then transferred to the cylinder.}
\end{center}
\end{figure}
\clearpage

\begin{figure}
\color{white}
\begin{center}
\subfigure[]{\includegraphics[%
  scale=0.3]{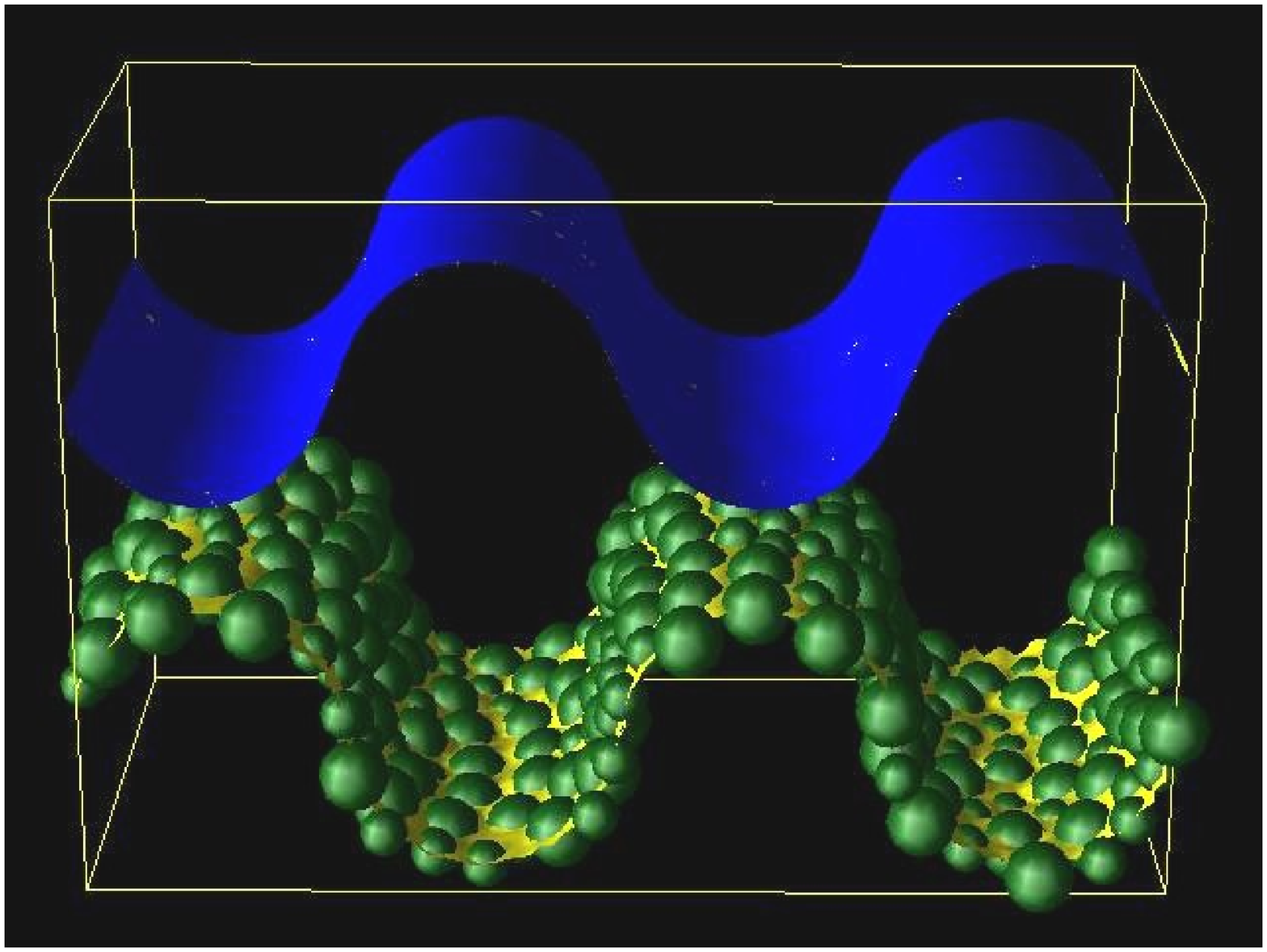}}\hskip0.05truein\subfigure[]{\includegraphics[%
  scale=0.3]{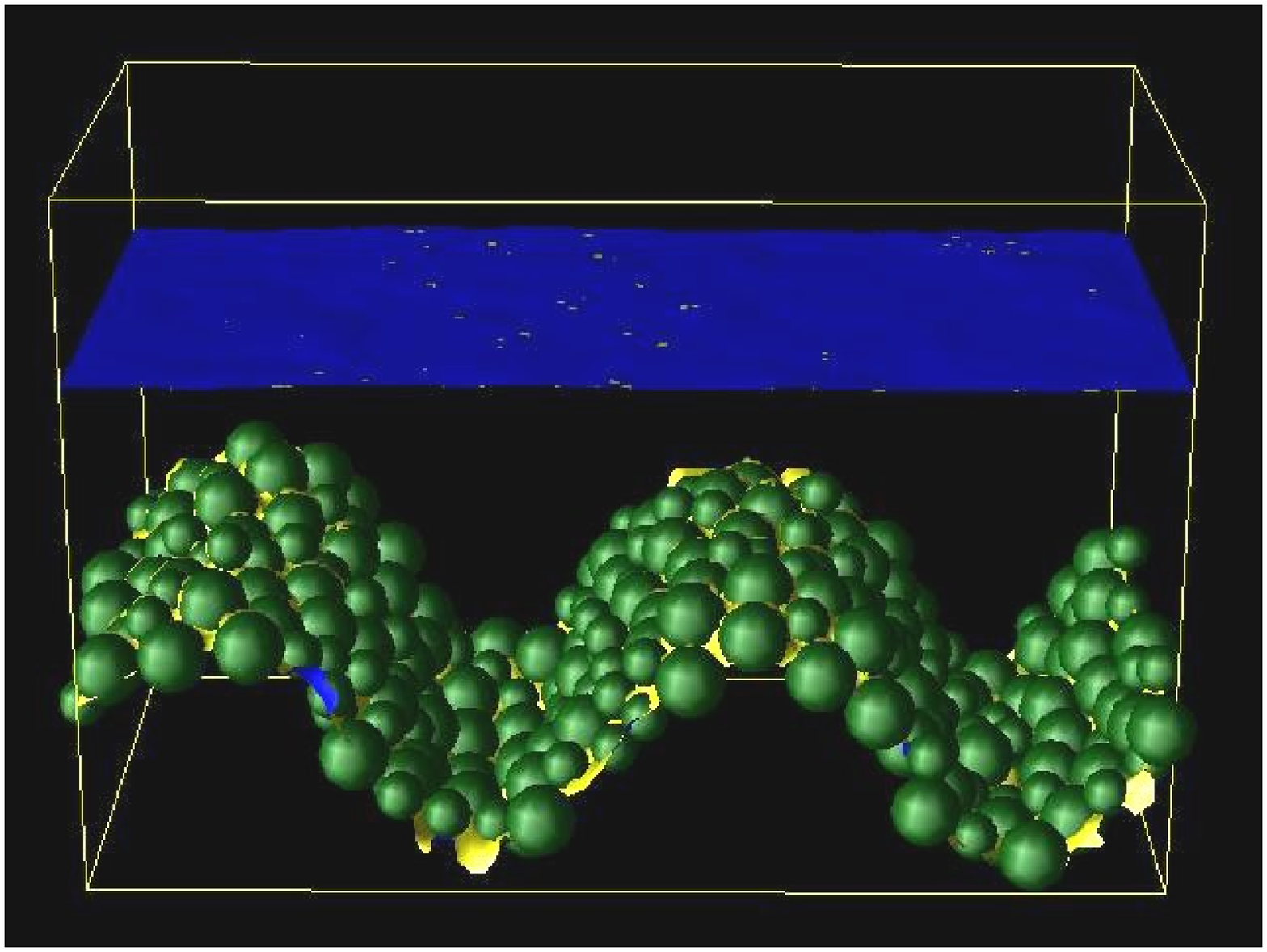}}
\\
\subfigure[]{\includegraphics[%
  scale=0.3]{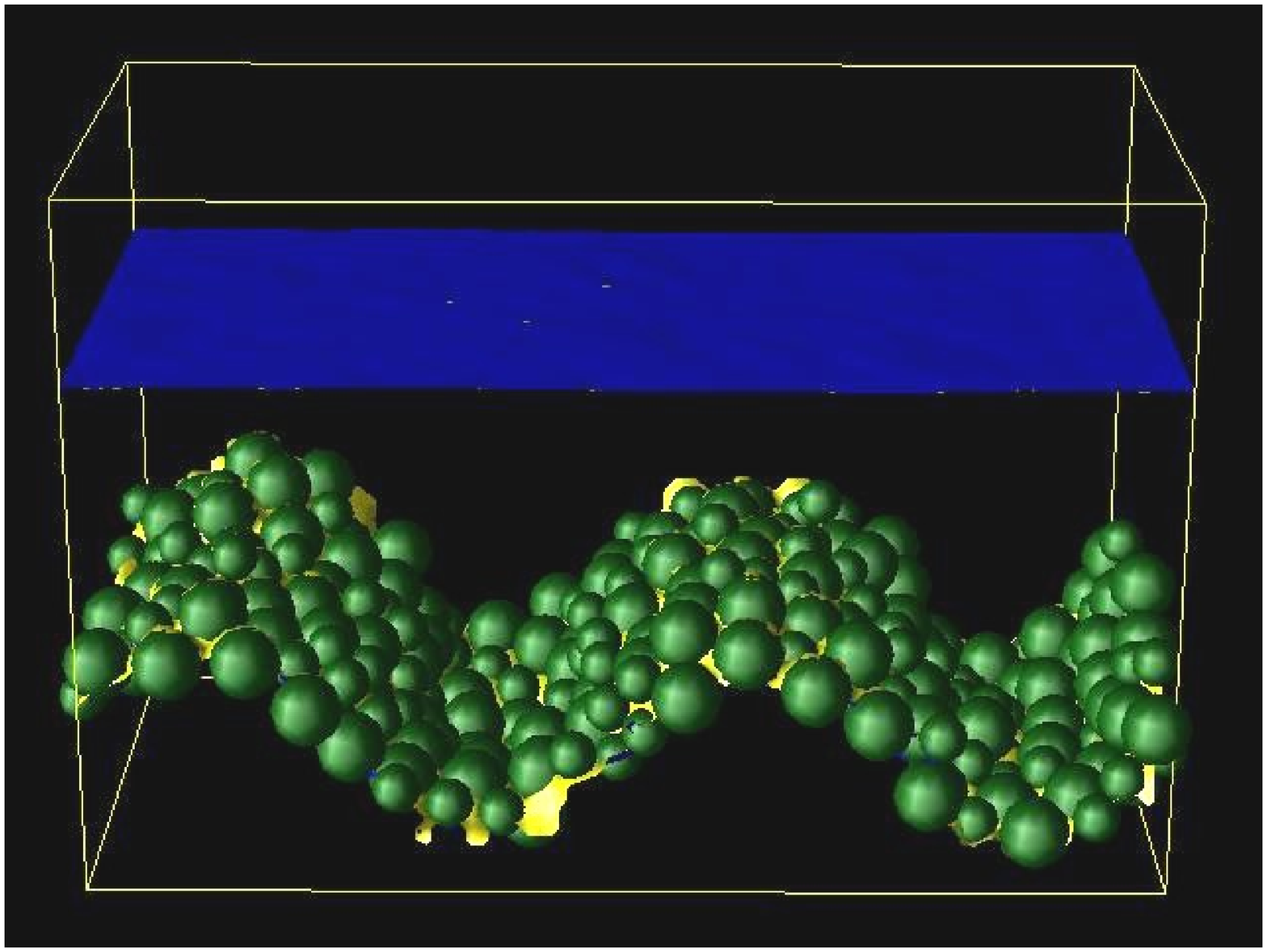}}
\subfigure[]{\includegraphics[%
  scale=0.3]{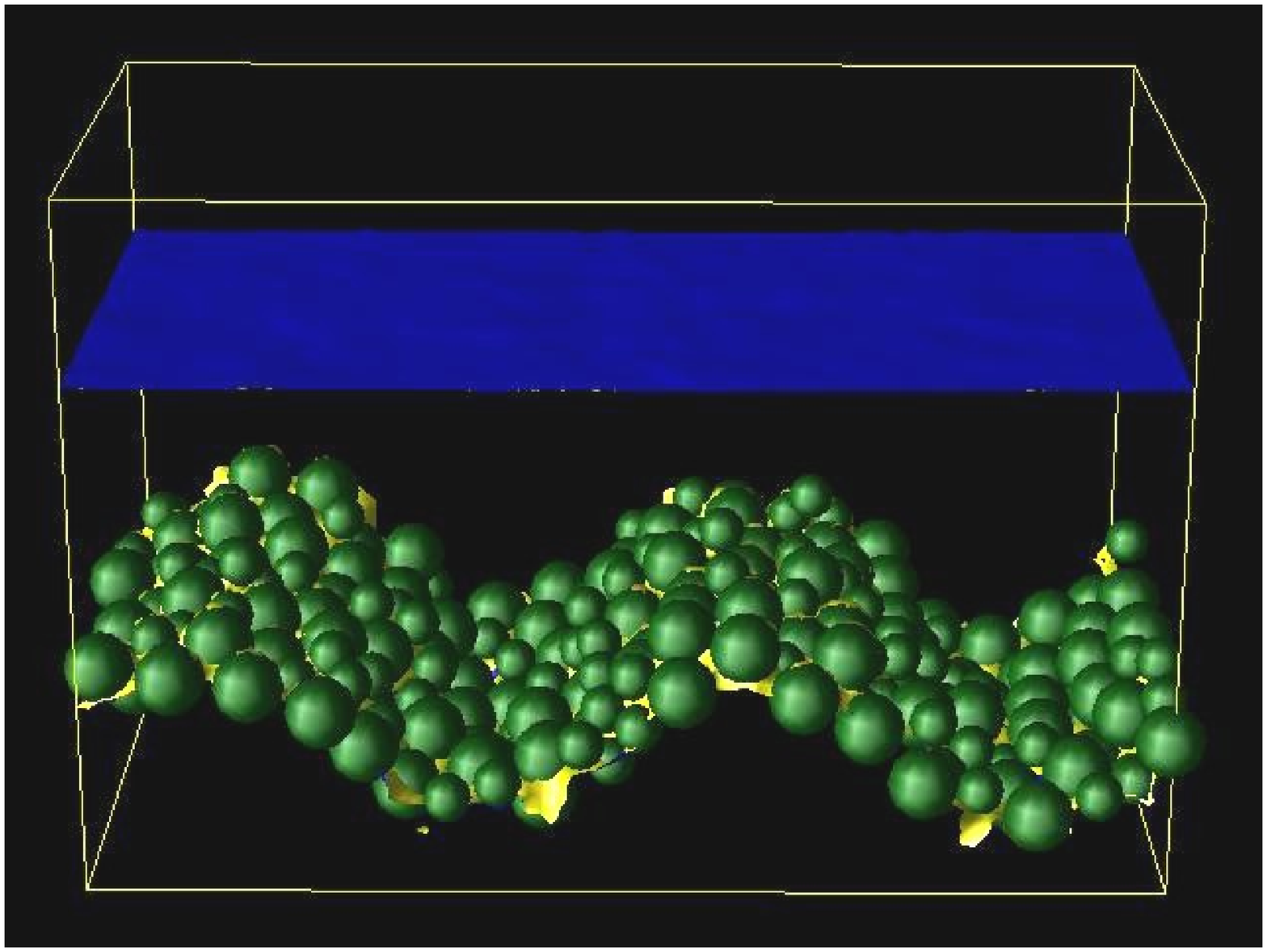}}
\caption{\color{black} {\bf Figure S2:} Time evolution of a periodically rippled interface coated with bidisperse, neutrally wetting colloids. The system occupies a $96\times 48\times 48$ lattice; particle radii are 2.1 and 3.2 lattice units. The initial state was created using a planar array of spheres generated by the algorithm of 
{\em (S10)}
on a flat periodic domain with coordinates then transferred to a repeating array of hemi-cylinders (upper left; after 1 000 timesteps). Subsequent frames (upper right, lower left, lower right) show 40 000, 80 000, and 150 000 timesteps. Almost no macroscopic motion of the jammed monolayer occurs after $t\simeq$ 80 000 timesteps. The evolution of a neighboring
uncoated fluid surface is also shown; this collapses to flatness roughly as $e^{-t/\tau}$ with $\tau \simeq$ 5 000 timesteps.}
\end{center}
\end{figure}

\clearpage

\subsection*{References for Supporting Material}

\noindent{\em (S1)}  V. M. Kendon, M. E. Cates, I. Pagonabarraga,
J.C. Desplat, P. Bladon, {\em J. Fluid Mech.} {\bf 440}, 147 (2001). 

\noindent{\em (S2)} K. Stratford, R. Adhikari, I. Pagonabarraga, J.-C. Desplat, {\em J. Stat. Phys.} in the press
(available at http://arxiv.org/abs/cond-mat/0407631).

\noindent{\em (S3)} N.-Q. Nguyen,  A. J. C. Ladd, {\em Phys. Rev. E}  {\bf 66}, 046708 (2002).

\noindent{\em (S4)} 
J.-C. Desplat, I. Pagonabarraga, P. Bladon,
{\em Comput. Phys. Comm.} {\bf 134}, 273 (2001).

\noindent{\em (S5)} R. Adhikari, K. Stratford, M. E. Cates, A. J. Wagner, {\em Europhys. Lett.} {\bf 71}, 437 (2005).

\noindent{\em (S6)} A. J. C. Ladd, {\em J. Fluid Mech.} {\bf 271}, 285 (1994).

\noindent{\em (S7)} E. D. Siggia, {\em Phys. Rev. A} {\bf 20}, 595 (1979).

\noindent{\em (S8)} M. E. Cates {\em et al.},
{\em J. Phys. Condens. Matt.} {\bf 16}, S3903 (2004).

\noindent{\em (S9)} S. A. Safran, {\em Statistical Thermodynamics of Surfaces, Interfaces, and Membranes}, Westview Press, Boulder (2003).

\noindent{\em (S10)}  A. Donev, S. Torquato, F. H. Stillinger, {\em J. Comput. Phys.} {\bf 202}, 737 (2005).


\begin{thebibliography}{}
\bibitem{ergodic} Note that the true equilibrium state of the system investigated here consists of two bulk fluid phases separated by a single flat interface; a thermodynamically negligible fraction of the colloidal particles densely coat this interface but the rest are spread through the fluid phases with the same uniform volume fraction as in the initial state. This minimizes interfacial area and maximises colloid entropy.
\bibitem{robinson1} There is an established technology for microreactors based on discrete droplet emulsions, but continuous operation is not possible unless all reagents and products are soluble in the continuous phase. See, e.g., \cite{robinson2}.
\end{thebibliography}
\end{document}